\documentclass[showpacs, nofootinbib]{revtex4}
\usepackage{amsmath,amssymb,amsthm}
\usepackage{graphicx}

\def\demi{\frac{1}{2} }

\newcommand{\dr}{\rightarrow}

\newcommand{\R}{\mathbb{R}}

\def\cc{{\cal C}}

\def\Ee{{\cal E}}

\def\oo{{\cal O}}
\def\ppp{{\cal P}}

\newcommand{\mone}{^{-1}}
\newcommand{\be}{\begin{equation}}
\newcommand{\ee}{\end{equation}}

\newcommand{\ket}[1]{|#1\rangle}

\def\la{\langle}
\def\ra{\rangle}
\newcommand{\one}{\mbox{$1 \hspace{-1.0mm}  {\bf l}$}}

\def\f{\frac}
\def\t{\textrm{Tr}}

\renewcommand{\v}{\overrightarrow}

\newcommand{\SO}{\mathrm{SO}}
\newcommand{\ISO}{\mathrm{ISO}}

\begin{document}
\title{Physics of Deformed Special Relativity:\\ Relativity Principle revisited}
\author{{\bf Florian Girelli\footnote{fgirelli@perimeterinstitute.ca}, Etera R. Livine\footnote{elivine@perimeterinstitute.ca}}}
\affiliation{Perimeter Institute, 31 Caroline Street North Waterloo, Ontario Canada N2L 2Y5}

\begin{abstract}
In many different ways, Deformed Special Relativity (DSR) has been argued to provide an effective limit of  quantum
gravity in almost-flat regime. Some experiments will soon be able to test some low energy effects of quantum gravity,
and DSR is a very promising candidate to describe these latter. Unfortunately DSR is up to now plagued by many
conceptual problems (in particular how it describes macroscopic objects) which forbids a definitive physical
interpretation and clear predictions. Here we propose a consistent framework to interpret DSR. We extend the principle
of relativity: the same way that Special Relativity showed us that the definition of a reference frame requires to
specify its speed, we show that DSR implies that we must also take into account its mass. We further advocate a
5-dimensional point of view on DSR physics and the extension of the kinematical symmetry from the Poincar\'e group to
the Poincar\'e-de Sitter group ($\ISO(4,1)$). This leads us to introduce the concept of a pentamomentum and to take
into account the renormalization of the DSR deformation parameter $\kappa$. This allows the resolution of the "soccer
ball problem" (definition of many-particle-states) and provides a physical interpretation of the non-commutativity and
non-associativity of the addition the relativistic quadrimomentum. In particular, the coproduct of the
$\kappa$-Poincar\'e algebra is interpreted as defining the law of change of reference frames and not the law of
scattering. This point of view places DSR as a theory, half-way between Special Relativity and General Relativity,
effectively implementing the Schwarzschild mass bound in a flat relativistic context.

\end{abstract}

\maketitle
\newpage

\tableofcontents
\newpage

\section*{Introduction}
Special Relativity will be centenary next year. It was introduced  in order to accommodate Maxwell's theory of
electromagnetism invariant under the Lorentz symmetry with the (Galilean) relativity principle. Among many features
Special Relativity  can be seen as the theory which changed the Galilean symmetries in order to have an universal
ratio: the speed of light $c$ is the same for all observers.

The appearance of this new maximum quantity puts into light two different regimes, the Galilean one, for speeds $v\ll
c$, and the relativistic one, for speeds $v\sim c$. The Galilean regime is still using the Galilean symmetry whereas
the relativistic  regime needs the shift to the Poincar\'e symmetry. Symmetries are naturally associated with the
physically relevant measured objects: either the Galilean momentum or the relativistic momentum. The notion of energy
is  of course also associated with the symmetries (more exactly to the Casimir invariant). Changing the symmetry meant
a shift to  a new concept of energy, and this is the well-known $E=mc^2$ for a particle at rest, opposed to $E=0$ in
the Galilean case.

Very soon after that Einstein introduced General Relativity, it became clear that one should look for a quantum version
of this theory.  If this is a real longstanding problem as still now unsolved, many arguments show that a minimal
length (or maximal mass or energy) should be a feature of this theory. These are usually dubbed Planck length, Planck
energy... By construction this minimal length should be universal, i.e. the same for any observer. This is obviously in
contradiction with the usual action of the Lorentz group. Two natural possibilities were envisaged in order to resolve
this issue either the Lorentz symmetry is broken or it is deformed. Recent experiments seem to indicate that the
Lorentz violation is not the one preferred by Nature, so the deformation seems to be the natural candidate to implement
the minimal length.

One of the possible ways to  deform Lorentz symmetry is done according to the quantum group  setting. The deformation
concerns mainly the action of the boosts on the translations, identified with the momenta. There are of course many
possible deformations, which physical meaning is quite unclear. There are other problems that plague this approach. The
biggest to our sense is that this deformation is made in a consistent way with the algebraic structure (coproduct)
describing the many body systems. Indeed if for example we are implementing a maximal mass, the Planck mass $M_P$, when
considering two bodies of masses smaller than $M_P$,  the 2-body system will still have a mass smaller than $M_P$. This
is a problem as of course macroscopic objects of mass bigger than the Planck mass can be thought made of particles of
mass smaller than the Planck mass. This is the so-called "soccer ball problem".
One can see therefore that if the  mathematical structure is clearly understood the physical understanding of this
"Deformed Special Relativity" (DSR) is not so clear.

This could be the an argument sufficient to discredit DSR as a relevant physical theory. There are however many
arguments apart from the implementing a minimal length, which tend to show that DSR is an important physical theory.
For example 3d gravity can be shown mathematically to be a DSR theory\cite{dsr3d}. There are also different, more or
less rigorous, arguments from quantum gravity theories which show that DSR is indeed a low energy description of a
"flat" quantum gravity \cite{dsrcosmo, dsrgol}. It is therefore really important  to understand the physical meaning of
DSR as it seems also to be an interesting candidate to provide a theoretical framework to the forthcoming Quantum Gravity
experiments like $\gamma$-ray bursts, GZK cutoff and so on \cite{amelino}.

\medskip
We intend here to provide such a general physical framework, which would solve all the different problems of physical
interpretation. We want for this to make  the analogy with the change of physics occurring in  the
Galilean-Relativistic change of symmetry  as re expressed in \cite{SR}. Indeed  we argued there that Special Relativity
could be seen as a deformation of the Galilean symmetry, in a very similar way to the deformation that occurs from SR
to DSR, see also \cite{chrissomalokos}. It seems therefore worth to push forward  the analogy. Let us summarize the
general situation.

We have now three physical regimes, Galilean, Relativistic, DSR. Let us consider a physical object $\oo$, where  $v$,
$L$, $M$ are its speed with respect to a chosen reference frame, its characteristic length (with respect to the same
reference frame), its mass (or energy). Note that General Relativity implies a non trivial relation between mass and
scale due to  the gravitational effect (Schwarzschild radius): there is a maximal mass $M_max\equiv \frac{c^2L}{G}$
associated to the scale $L$. We can then distinguish the different regimes. In the Galilean regime, $v\ll c$. In the
relativistic regime, we have $v \sim c$ and $M\ll M_{max}$. The DSR regime is defined as $M\sim M_{max}$. First note
that relevant DSR effects still happen when $v\sim c$. But more relevant to our discussion, the DSR regime is naturally
reached near the Planck scale when quantum effects (Compton length) induce mass fluctuations of the same order of
magnitude as the maximal mass $M_{min}\equiv\frac{\hbar}{c L} \sim M_{max}$. Associated with the different physical
regimes, one has different symmetries. One starts with the Galilean group, and deforms them into the Lorentz group and
the Poincar\'e group. Physically we also have a shift of the notion of momentum and energy from Galilean quantities to
relativistic ones.

We want to argue that the same occur when going from SR to DSR. We shift into a new symmetry, the de Sitter group (or
anti de Sitter) naturally extended to the Poincar\'e de Sitter group. The new physical objects are associated with this
symmetry (Poincar\'e de Sitter) and we have therefore a new momentum (pentamomentum) and a new notion of energy (DSR
energy). Note that the de Sitter group is bigger than the Lorentz group, which can be explained by the fact the
symmetry takes also into account the characteristic scale so that it is preserved for any observer. The pentamomentum
naturally encodes the renormalization of the $\kappa$ factor,  so  in this sense we solved the "soccer ball problem".
Just as in the relativistic case we had a shift in the understanding of space-time, there is also a new notion of
space-time, actually two equivalent ones. Either one can consider it as fuzzy with a scale encoding the precision, or
as a space-time with two times, one external and the other internal to the object (proper time). It is then natural to
see a shift in the notion of simultaneity and the relativity principle. Einstein introduced the relative simultaneity,
opposed to the absolute galilean one.

If there is a new symmetry, one has still to interpret the physical meaning of the coproduct that was thought as
representing the multiparticles state. In the relativistic regime, a particle is described by its speed relative to a
chosen reference frame, and under change of reference frames, only the speeds mattered. In the DSR regime, the masses
of the reference frames have also to be taken into account\footnote{Note that the importance of the notion of reference
frame in  DSR has also been emphasized recently in \cite{liberati, toller}}. This explains the problems of non
commutativity, or non associativity that made the physical interpretation of the coproduct difficult. The key points
is that multiparticles states and the law of scattering are described in terms of the momentum associated to the
symmetry of the physical regime, and that one also should describe the law of change of reference frames (which are in
general non-commutative and non-associative). Moreover these different structures must of course be compatible with
each other.

We are going to develop in details the DSR regime, and show clearly how the mentioned problems are solved.
Before, let us summarize the general philosophy  in the following table.
\begin{equation}\nonumber
\begin{array}{|c|c|c|}
\hline \textrm{Galilean Relativity}  & \textrm{Special Relativity} & \textrm{Deformed Special Relativity} \\
\hline (\overrightarrow{v},\overrightarrow{p_g}) &(\overrightarrow{v}, \gamma \overrightarrow{v}, p^{\mu}=
\left(\begin{array}{c}\gamma \\ \gamma \overrightarrow{p_g}\end{array}\right))& (\overrightarrow{v}, \gamma
\overrightarrow{v}, p^{\mu}, \pi^A=\left(\begin{array}{c} c\kappa\Gamma \\ \Gamma
p^{\mu}\end{array}\right))\\
\hline \textrm{Absolute space, time:}\;(\overrightarrow{x},t) & \textrm{Space-time:}\; x^{\mu} & \begin{array}{c}
\textrm{Space-time with
internal time: }\; (x^{\mu}, T)\\ \textrm{or fuzzy space-time with resolution:}\; (X^{\mu}, \kappa^{-1})\end{array} \\
\hline \textrm{Absolute reference frame}& \begin{array}{c} \textrm{Relative reference frame,} \\  \textrm{description
by relative speed}\end{array} &
\begin{array}{c}  \textrm{Relative reference frame,} \\ \textrm{description by relative speed, relative momentum}\end{array}\\
\hline \textrm{Absolute simultaneity  }        &  \textrm{Relative simultaneity  } &\begin{array}{c} \textrm{Relative
simultaneities from internal and external times}\\ \textrm{or fuzzy simultaneity}\end{array} \\
\hline \textrm{No bound  } & \begin{array}{c} v, v_1\oplus v_2 \leq c \\ \textrm{No bound on }\; \gamma v, \; p^{\mu}\end{array} & \begin{array}{c} v, v_1\oplus v_2 \leq c \\
                                    p^{\mu}, p^{\mu}_1\oplus p^{\mu}_2 \leq
                                                                        \kappa \\\textrm{No bound on }\; \pi^A
                                                                        \end{array}\\ \hline
\end{array}
\label{eq:table}
\end{equation}
For clarification, let us say that  we used the Snyder coordinates to define $\pi^A$, and we have $\gamma^2=
\frac{1}{1-\frac{1}{c^2}v^iv_i}$ the usual relativistic factor in SR, while
$\Gamma^2=\frac{1}{1-\frac{1}{(c\kappa)^2}p^{\mu}p_{\mu}}$ is its analog for the DSR regime (Snyder coordinates).

\medskip

To have a clear heuristic description of the new physical regime,  we shall first recall some basic facts about the
different maximal or minimal quantities one can introduce. We then move on to recall the basic definition of DSR and
the problems which are met in this context. We introduce then the DSR momentum, with the new symmetry and take one by
one the different problems to show how they are solved within this context. We expose then some of the new physical
features.

\section{Motivating DSR: Minimal-Maximal Quantities} \label{minmax}

In General Relativity, considering a region of length scale $L$, we  can bound the mass of the content of that region
by the mass of the Schwarzschild black hole of radius $L$: \be M_L= \frac{c^2L}{G} \ee where $G$ is Newton's constant.
Note that the true formula should contain a factor $1/2$ on the right hand side, which we forget in the following as
long as we work only with orders of magnitude. $M_L$ can be interpreted as defining a maximal energy density for a
phenomenon of length scale $L$. Note that for a particle at rest, bounding the mass is equivalent to bound its energy.

On the other hand, quantum mechanics allows to define a notion of quanta of mass linked to a phenomenon of length scale
$L$: \be \delta M = \frac{\hbar}{cL} \ee where $\hbar$ is the unit of action introduced in the quantum theory.  $\delta
M$ can be interpreted as a minimal mass of any phenomenon of length scale $L$, or as the uncertainty or fluctuations in
the mass of such a phenomenon. This mass is usually very small for usual objects (the electron for example).

Dealing both with general relativity and quantum mechanics, we then get both a minimal and maximal bound on masses associated to a length scale $L$:
\be
\frac{\hbar}{c L}
\equiv
M_{min}
\leq M\leq
M_{max}
\equiv
\frac{c^2L}{G}.
\ee

The Planck scale is then defined as the scale at which the mass fluctuations $\delta M$ become comparable to the
maximal mass: \be M_{min}=M_{max} \,\Rightarrow\, L\equiv L_P=\sqrt{\frac{G\hbar}{c^3}}. \ee The corresponding mass is
called the Planck mass $M_P=\sqrt{\frac{c\hbar}{G}}$. The Planck scale is usually  considered as indicating a new
regime - the quantum gravity regime.

The motivation to postulate a maximal mass or energy in DSR is to take into account  this maximal bound on mass
existing in general relativity. Moreover when the mass $M$ is close to the bound $M_L$, we expect strong physical
effects  (as described by General Relativity for example) and thus modifications of the Special Relativity
phenomenology. In this new regime, dubbed DSR regime, it is clear that the mass of the system becomes relevant to its
kinematics/dynamics (contrary to the Special Relativity set-up). A natural consequence is that we expect an extension
of the relativity principle in DSR which takes into account the mass of reference frames.

The traditional view is to postulate a universal maximal mass/energy being the Planck
mass, which could be measured by every observer. Our point of view is that it does make sense to postulate the
universality of the Planck mass -as a signature of the quantum gravity regime- but it doesn't make sense to postulate
it is a bound on energy/mass. Indeed, macroscopic objects have rest energies much superior to the Planck mass. They
would hardly make sense in a theory bounded by $M_P$. This paradox is usually referred in the DSR literature as the
{\it soccer ball problem}. What appears in our simple presentation is that a DSR theory should naturally include a
description of the renormalization of scales, and thus explain how the energy/mass bound get renormalized with the
scale $L$. Assuming that general relativity is exactly valid down to the Planck scale, we expect that the mass bound
get resized linearly with the length scale $L$ as expressed above.

These considerations also allow to explain why it is natural to expect that the bare speed of light  $c$ -at long
wavelength- gets renormalized in DSR theories. Indeed let us consider a light wave of wavelength $L$. To study its
propagation, it is enough to cut the space(time) into regions of size $L$. Then if the energy/mass content of each
region is saturated to the maximal mass $M_L$, this means that the space(time) is made of black holes of size $L$ which
would trap the light and forbid it from propagating: the speed of light would vanish. Forgetting about a fundamental
vacuum energy (the cosmological constant) and the own energy carried by the light wave, energy/mass fluctuations in
regions of size $L$ are of order $\delta M$. As long as $\delta M\ll M_L$, there is no significant change of the metric
of space-time and thus no need to renormalized the speed of light. But as soon as $\delta M$ becomes comparable to
$M_L$, we need to take into account the non-trivial metric induced by the energy/content of each space-time cell, which
would modify the speed of light \cite{vsl}. The extreme case is then the Planck scale, at which energy/mass quantum
fluctuations would saturate "automatically" the mass bound: the speed of light $c(L=L_P)$ would vanish.

\medskip

The notion of quanta of mass and the procedure of cutting up the space(time) into cells of given size prompt the idea
of {\it binding energy}.  When dealing with atomic structure, the mass of the total system is either bigger or smaller
than the sum of the mass of its constituents, depending of the interaction between the constituents.
Now assume that we are considering two objects of rest energy or mass $M$ distant of length $L$. Then in the newtonian
approximation, the  two objects will see the potential
$$
V(1,2)= -\frac{GM^2}{L},
$$
where $G$ is the Newton constant. In the case that the mass $M$ is actually the mass bound $M_L$, then the interaction
energy is exactly $V(1,2)=-M_Lc ^2$, so that the total energy of the composite system would be $E_{tot}=2M_L c^2-M_L
c^2=M_L c^2$: this procedure of putting two objects saturating in the mass bound at the actual considered length scale
leaves the maximal mass $M_L$ invariant. Applying this argument to the Planck mass $M_P$ can actually be thought as the
reason to assume in DSR that the Planck mass to be a universal maximal mass. But this is only a hand-wavy  argument
that one can not fully trust. Indeed, one must first assume the validity of the Newtonian approximation out of its
known domain of validity. Moreover two objects of size $L$, and then assumed to saturate the mass bound $M_L$, can not
be put at a distance smaller than $2L$ between each other. If we were putting them at a distance $L$, they then should
be considered as forming a single object, and we should modify the mass bound to take into account the input from
general relativity. If they are at a distance larger than $L$, then the interaction energy will be smaller in absolute
value and thus the total (rest) energy of the system will be bigger than $M_L$.

Nevertheless, we believe that the concept of the (gravitational) {\it binding energy} should be a key point in DSR
theories.  Indeed we expect that DSR deforms the notion of energy of Special Relativity in a similar way that Special
Relativity deformed the Galilean and Newtonian notion of energy \cite{SR}. This new notion of energy should take into
account some gravitational effects, as it is believed it describes an effective theory for quantum gravity in a flat
background space-time (see for example \cite{dsrgol, dsrcosmo}). In the following sections, we will attempt to describe
in more details how the energy of a free system in a DSR theory can be seen as containing an interaction energy term
when compared to the notion of energy of Special Relativity. Such a notion of DSR energy is essential in order to
understand the behavior of macroscopic objects.

Introducing bounds on mass/energy is not consistent with Special Relativity.  More precisely the standard action of
boosts doesn't respect a bound in energy\footnote{The case of a bound on possible masses of particles is more subtle.
In special relativity, there is simply no reason to have a minimal or maximal rest mass. One could either put it by
hand, although it would be more satisfying to derive such a bound from a symmetry principle. Or one could modify the
measure on the momentum space to take into account that bound, in which case one clearly modifies the symmetries of the
momentum space.}. Therefore to include such a bound, we need to modify this kinematical theory, by deforming the
Lorentz symmetry group or at least its action on the configuration space of the theory. This is exactly the same
approach as when we modified the Galilean framework to accommodate a universal maximal speed and thus obtain Special
Relativity. The main difference is that we had experimental evidence for a universal speed of light, while DSR has been
created out of a theoretical idea that an effective framework for a quantum gravity theory merging general relativity
with quantum mechanics on a almost flat space-time should lead to such a mass/energy universal bound. Nevertheless,
experiments will soon be launched (GLAST, ..) and we will be able to test the predictions of DSR.


\section{Deformed Special Relativity : features and problems}
In this section we recall the main features and problems of the DSR. Note that this construction is very similar to the
construction of Special Relativity from a deformation of the  Galilean point of view \cite{SR}. We shall see later that
taking this analogy seriously allows to solve most of the problems. For the most recent update on DSR see the lectures
by Kowalski-Glikman, or Amelino-Camelia \cite{kg:lecture, amelino}.

\subsection{Features}
Let us recall now the features of the Deformed Special Relativity (also coined Doubly Special Relativity). The first
example of such theory is pretty old and a well known example of non commutative geometry. Snyder in 1947 introduced a
theory which would naturally incorporate a cut-off in the momentum space, without breaking  the symmetries, in order to
regularize field theory \cite{snyder}.

He showed that by starting with a non trivial momenta space, a de Sitter space, one could keep the symmetries fine, at
the price of getting a non commutative space-time. He considers the momentum as an element of the space $SO(4,
1)/SO(3,1)$, which can be parameterized, using the  five dimensional Minkowski space coordinates\footnote{We note the
5d indices with capitals, the 4d indices with greek letters, the 3d indices with latin indices.} $\pi_A$,
\begin{equation}\label{desitter}
-\kappa^2= +\pi_0^2-(\pi^2_1+\pi^2_2+\pi^2_3+\pi^2_4)=\pi^{\mu}\pi_{\mu}-\pi^2_4.
\end{equation}
$\kappa$ is a constant with dimension of a mass,   $\pi_4$ is the conserved direction and we can therefore see that we
kept the Lorentz symmetry as the subgroup and the four left transformations, called the de Sitter boosts are identified
with the momentum. The Lorentz part $J_{\mu\nu}$ (c.f. appendix for the notations) of $SO(4,1)$
 is acting in the regular way on the Minkowski coordinates $\pi_{A}$,
\begin{equation}\label{desitter1}
\begin{array}{rcl}
[M_i, \pi_j]&=& i\epsilon_{ijk}\pi_k,  \; [M_i, \pi_0]=[M_i, \pi_4]=0, \\
{[}N_i, \pi_j{]}&=& \delta_{ij}\pi_0,  \; [N_i, \pi_0]= i\pi_i,  \; [N_i, \pi_4]=0,
\end{array}
\end{equation}
where we respectively noted as usual $M_i=\epsilon_{ijk}J_{jk}$, $N_{i}= J_{0i}$ the rotations and the boosts.

Essentially this deformation of the momentum space can be understood as a map from $\R^4$ to de Sitter. This
deformation can be compared with  Special Relativity (SR) arising as a deformation of the space of speeds: there,
$\R^3$, the space of speeds is sent to the hyperboloid $SO(3,1)/SO(3)$. There is therefore a strong analogy between the
SR case and the Snyder approach.

Space-time is now seen as the tangent space of the de Sitter space and the coordinates are therefore given by the (non
commuting) de Sitter boosts generators.
\begin{equation}
\begin{array}{rcl}
(X_{i}, X_0)&=& i \frac{\hbar}{c\kappa}( J_{4i},\frac{1}{c}J_{40}),\\
{[}X_{i}, X_{j}{]}&=& i  \frac{\hbar^2}{(c\kappa)^2}J_{ij},
\end{array}
\end{equation}
and so on. On the other hand, the momentum is given by a  coordinate system on de Sitter and Snyder chooses to take as
an example $p^{\mu}=c\kappa \frac{\pi^{\mu}}{\pi^4}$, which generates a deformed commutators between  position and
momentum.
\begin{equation}
\begin{array}{rcl}
[X_{i}, p_{j}]&=& i \hbar (\delta_{ij} +\frac{1}{(c\kappa)^2}p_{i}p_{j}),
\end{array}\end{equation}
and so on.
The topic of implementing a maximum quantity in a compatible way was then left aside for many years until the new
arising phenomenology of both Quantum Gravity and String theory prompted a new interest on the subject, in particular
first from the quantum group point of view \cite{ruegg} and then from the more phenomenological point of view
\cite{amelino1}.

It was of interests for the algebraists as indeed introducing a maximal quantity to be respected from the Poincar\'e
symmetry consists in deforming the boosts action. In general one can introduce the Poincar\'e Lie algebra and then keep
the Lorentz part non deformed,
\begin{equation}\label{lorentz}
\begin{array}{rcl}
[M_i, M_j]&=& i\epsilon_{ijk}M_k, \;[N_i, N_j]= -i\epsilon_{ijk}M_k, \; [M_i, N_j]= i\epsilon_{ijk}N_k, \\
{[}M_i, p_j{]}&=& i\epsilon_{ijk}p_k, \; [M_i, p_0]= 0,
\end{array}
\end{equation}
whereas the boost action on the momenta (identified with the translations) is deformed.
The most general deformation is given by
\begin{equation}\label{deformation1}
\begin{array}{rcl}
[N_i, p_j]&=&A\delta_{ij}+B p_ip_j + C \epsilon_{ijk}p_k,   \\
{[}N_i, p_0{]}&=& D p_i,
\end{array}
\end{equation}
where $A, B, C, D$ are functions of $p_0, p_i^2, \kappa$. We would like that the deformed Poincar\'e group becomes the
usual Poincar\'e group in the continuum limit where $\kappa\rightarrow \infty$. This gives therefore some conditions on
these functions ($A, D \rightarrow 1$, $B, C\rightarrow 0$). We can moreover show that the function $C$ has to be zero
from the Jacobi identity, and also we must have the differential equation
\begin{equation}\label{deformation2}
\frac{\partial A}{\partial p_0}D+ 2\frac{\partial A}{\partial \overrightarrow{p}^2}(A+ \overrightarrow{p}^2B)-AB=1.
\end{equation}
Different solutions of this equation, with the limit conditions for $\kappa\rightarrow\infty$ give us different
deformations. The most used one are coined DSR1, DSR2, denominations which varies according to the authors. More
specifically, we can mention the bicrossproduct basis,  the Magueijo-Smolin basis. Note that different deformations
corresponds to different physical situations, i.e. different  quantities bounded in a covariant way.

Kowalski-Glikman noticed then that in fact all these algebraic  deformations could be geometrically understood: they
are different coordinates system on the de Sitter space \cite{kgsnyder}. This important remark    put, at the same
time, Snyder's approach back in the game as a DSR candidate. More precisely, when restricting the $\pi$'s to the
homogenous space $SO(4, 1)/SO(3,1)$, in a particular coordinates system, e.g. $\pi_0= \pi_0(p_0, \overrightarrow{p}),
\pi_i = p_i \pi_4(\pi_0(p_0, \overrightarrow{p})), \pi_4 = \sqrt{\kappa^2+ \sum_{i=0}^{3}\pi_i^2 }$,  one then recovers
the previous commutation relations (\ref{lorentz}, \ref{deformation1}), with the functions $A,B,D$  expressed in terms
of $\pi_A$.

For example if one sets
\begin{equation}
\begin{array}{rcl}
\pi_0&=&-\kappa \sinh\frac{p_0}{\kappa}-\frac{\overrightarrow{p}^2}{2\kappa}e^{\frac{p_0}{\kappa}}\\
\pi_i&=&-p_ie^{\frac{p_0}{\kappa}} \\
\pi_4&=& -\kappa \cosh\frac{p_0}{\kappa}-\frac{\overrightarrow{p}^2}{2\kappa}e^{\frac{p_0}{\kappa}},
\end{array}
\end{equation}
one gets the bicrossproduct basis  (in order to avoid confusion we note these  coordinates  $P_{\mu}$) \cite{ruegg},
\begin{equation} \label{bicross1}
\begin{array}{rcl}
[N_i,P_j]&=&i\delta_{ij}(\frac{\kappa}{2}(1-e^{\frac{-2P_0}{\kappa}})+\frac{1}{2\kappa}\overrightarrow{P}^2)
-i\frac{1}{\kappa}P_iP_j\\
{[}N_i,P_0{]}&=&iP_i
\end{array}
\end{equation}
where the 3-d momentum  is bounded, whereas the energy can become infinite. From the commutation relations
(\ref{desitter1}), one sees that $\pi_4$ can be interpreted as the Casimir of the deformed symmetry. We therefore have
the associated Casimir in this basis given by
\begin{equation} \label{bicross2}
\begin{array}{rcl}\cc= \kappa^2\cosh\frac{P_0}{\kappa}-\frac{\overrightarrow{P}^2}{2}.\end{array}
\end{equation}

The other important example is  the Magueijo-Smolin basis (with coordinates noted $\ppp_{\mu}$), which corresponds in
this case to the coordinates
\begin{equation}
\pi_{\mu}= \frac{\ppp_{\mu}}{1-\ppp_0/\kappa}.
\end{equation}
and the deformation is then
\begin{equation}
\begin{array}{rcl}
[N_i,\ppp_j]&=& i(\delta_{ij}-\frac{1}{\kappa}\ppp_i\ppp_j) \\
{[}N_i, \ppp_0{]}&=&i(1-\frac{\ppp_0}{\kappa})\ppp_i.
\end{array}
\end{equation}
where both the energy and the 3d momentum  are bounded.  The associated Casimir in this basis is given by
\begin{equation} \label{magsmol2}
\begin{array}{rcl}\cc= \frac{\ppp_0^2-\overrightarrow{\ppp}^2}{(1-\frac{\ppp_0}{\kappa})^2}.\end{array}
\end{equation}

It is important to remember that the bounds that are introduced according to the chosen DSR are covariant, i.e.
preserved under the Lorentz transformations, and this by construction. There is no breaking of the Lorentz symmetry.

One can now take advantage of the algebraic structure and define the space-time structure underlying this momentum
space. This is done through the so-called Heisenberg double. This technic takes advantage from the fact that the phase
space can be interpreted as a cross product algebra: a pair of dual Hopf algebra, acting over each other. One is
representing the momenta algebra, whereas the other one is the space-time algebra. They are acting over each other by
translations.

One needs therefore to make the momenta algebra a Hopf algebra, and so to define the coproduct $\Delta$ by duality from
the product. It is clear that the shape of the coproduct will depend as well on the choice on the functions $A,B,D$. It
is then used to define the product on the space-time coordinates, as there are defined as dual to the momenta.
\begin{equation}
\begin{array}{rcl}
<p_{\mu}, x_{\nu}>&=&g_{\mu\nu}, \; \; \textrm{with } g_{\mu\nu}= \textrm{diag}(-1,1,1,1) \\
<p, x_1.x_2> &=& <\Delta p,x_1\otimes x_2 >.
\end{array}
\end{equation}
The result is a non commutative space, which has a different non commutativity according to the chosen momentum
deformation.  The usual pinpointed coordinates are the $\kappa$-Minkowski coordinates $x$, which can be related by an
adequate change of coordinates to the Snyder coordinates:
\begin{equation}
\begin{array}{rcl}
[x_0, x_i]&=& -\frac{i}{\kappa}x_i, \; \; [x_i, x_j]= 0,\\
{[}X_{\mu}, X_{\nu}{]}&=& -i\frac{\hbar^2}{c^2\kappa^2}J_{\mu\nu}.
\end{array}
\end{equation}

 Initially the theory started as a mathematical trick to regularize Quantum Field Theory, having in mind  a possible manifestation of
Quantum Gravity. This point got more and more sustainable as for example when considering the 3d quantum gravity case,
one can explicitly show that the algebra of observables for one particle is given by a DSR algebra \cite{dsr3d}. Other
arguments showed that the effective description of the 4d QG case should also be given by a DSR algebra
\cite{dsrcosmo,dsrgol}. In this sense, the new quest for the QG phenomenology has to go through some DSR
considerations.

To understand the physical implications of this effective theory is therefore very important. However if the theory, on
its mathematical aspects, is more or less understood, the underlying physics is much less because of some strange
features that are arising. This problem of interpretation is drawing more and more attention and there is an increase
on the number of different propositions to solve this  problem. 
Another
manifestation of this problem is for example the notion of speeds in the DSR regime. Many articles, and propositions
can be found on this topic.

To our mind it is important to define first a general physical setting and then check these problems. Hopefully by
finding the good one, all the problems should clear out. We intend to provide such a general scheme but before let us
pinpoint the main problems one encounters in DSR.

\subsection{Problems}\label{pbs}
In this section we enumerate and give the most important problems which are accompanying the usual DSR interpretations.

\subsubsection{Multitude of deformations}\label{pbdeformation}
The first question which can come up when looking at the definition of DSR, is that "are all these deformations
physically equivalent?" Meaning: is there one that should be pointed out by physics or somehow Nature doesn't make the
difference between them? This is essential as a priori each deformation corresponds through its Casimir to some new
dispersion relations, and conservation laws.

From the algebraic point of view it seems that only one deformation is physical, and experiments should pick up only
one. If however one is more inclined to the geometric setting then one usually thinks that all coordinate systems are
equivalent and in this way, all the deformations are equivalent.

Of course the two approaches clash and  the answer should lie in a better formulation of the arguments, which we shall
do later.

\subsubsection{Coproduct: non-commutativity (spectator problem) and non-associativity?}\label{pbnc}
To add the momenta, i.e. to deal with a many body system one usually considers the coproduct associated to the algebra
of symmetries. For example in the two particles case (undeformed Poincar\'e), the scattering is described by the
trivial coproduct
$$
\Delta P= 1\otimes P + P\otimes 1,
$$
which applied on a two particles states $\ket{1,2}$ gives the usual addition $p^{(1)}+p^{(2)}$. For most of the DSR
basis, the associated coproduct is not symmetric in this way (cocommutative), which brings then the  "spectator
problem".

As an example let us recall the coproduct for the bicrossproduct basis:
\begin{equation}\begin{array}{rclcrcl}
\Delta p_0&=&1\otimes p_0+p_0 \otimes 1 &\rightarrow& p_0^{(1+2)}&=& p_0^{(1)}+p_0^{(2)}\\
\Delta p_k&=&p_k\otimes e^{-\frac{p_0}{\kappa}}+1\otimes p_k&\rightarrow& p_k^{(1+2)}&=&
p_k^{(1)}e^{-\frac{p_0^{(2)}}{\kappa}}+p_0^{(2)}.
\end{array}
\end{equation}
It is then obvious that the order in which we label the particles is then relevant, which is the manifestation of the
non-associativity and then one can ask about the rest of the Universe, or some spectators looking at the scattering.
\begin{equation}
\begin{array}{rcl}
p_0^{(1+2+univ)}&=& p_0^{(1)}+p_0^{(2)}+ p_0^{univ}\\
p_k^{(1+2+univ)}&=& (p_k^{(1)}e^{-\frac{p_0^{(2)}}{\kappa}}+p_0^{(2)}) e^{-\frac{p_0^{(univ)}}{\kappa}}+ p_k^{(univ)}.
\end{array}
\end{equation}
This shows that we have a non local interaction, which is that the rest of the universe is contributing for any
scattering, for all time and positions, independently of the separation of the two particles \cite{kg:lecture}. This is
pretty much against the usual physical intuition.

Note that when $p_0\dr \infty$, there is no issue anymore, this corresponds to an infinite energy, or to an infinite
mass of the mass reference frame. This remark will take all of its importance later on.

There exists also some versions of the coproduct which is commutative but is not coassociative. Of course they don't
correspond to quantum group like deformations. As an example one can cite the proposal by Magueijo-Smolin, tentative to
solve the soccer ball problem. We come back on this in the following subsection. Another example is associated to the
Snyder like deformation. It is not difficult to see that then the addition of momentum is both non commutative and non
associative in general. This case hasn't been studied intensively in the existing litterature\footnote{Except in
\cite{kgsnyder}, who however didn't mention these features.}, but we come back to this point in the section
\ref{pheno}. Note that this is very similar to the addition of speeds in the SR case.

\subsubsection{The soccer ball problem}\label{pbsoccer}
The goal to DSR is to introduce a bound which is compatible with a deformation of the symmetry. Now if one takes the
point of view that the coproduct of the algebra gives the scattering rule, one sees that we are too much successful: if
indeed the algebra describing one particle contains in a consistent way the bound, the two particle state will also
have the same bound. This is not good as one expect to see the usual Special Relativity at our scale: Kowalski-Glikman
coined this problem {\it  the soccer ball problem} as if the quanta making the soccer ball can have energy bounded by
the Planck energy $E_P$, the soccer ball has a much bigger energy than $E_P$. The coproduct doesn't seem then to give
the right notion of scattering, and from the construction, cannot.

This can be compared to the SR case where speeds are constrained to live on the hyperboloid. The addition of speeds is
given by the vectorial addition on the hyperboloid and the sum of two speeds still lives on the hyperboloid, and so is
still bounded by the speed of light as expected.

  Right in the roots of the DSR philosophy seems to be a
problem: how do we get the classical world, which does not possesses any bound on the energy or mass?

Some solutions have been proposed to solve this problem and we would like to present the one by Magueijo-Smolin
\cite{leejoao}. It naturally followed from their approach to DSR, namely as a deformation of the representation of the
Poincar\'e group. In order to add momenta they propose to invert this map to get the linear momenta, add them in this
space and then resend the result in the deformed space. Let $f:\R^4\dr dS_{\kappa}$ be the deformation then
$$
p_{tot}=f(f\mone(p_1)+f\mone(p_2)).
$$
This proposition for addition is not solving the soccer problem, but if one adds a characteristic unity $\lambda$ for
each particle, and then rewrite the previous addition  as
$$
p_{tot}=f_{2\lambda}(f_{\lambda}\mone(p_1)+f_{\lambda}\mone(p_2)),
$$
one can see that this is solving the problem. If the intuitive origin of this proposition is clear -a renormalization
group picture- it is not at clear to understand the mathematical setting of this proposition: first it seems to be non
associative, and then doesn't come from any usual mathematical deformation, encountered in the quantum group context
(which respects the associativity or co-associativity).

To summarize, if very attractive as a possible solution, it has a lot to be understood.

We shall see in the next section that by introducing a new notion of momentum, as new as the relativistic momentum was
with respect to the galilean momentum,  we can solve all these problems and at the same time give a precise context and
explanation to the Magueijo-Smolin proposition.

\section{Solving the problems: the pentamomentum} \label{pentamomentum}
In this section, we want to propose a general scheme which will solve the problems we mentioned before.

Essentially being in a new regime, we have to change of symmetry. This is what was in done in the context SR, and as by
changing the symmetries we are able to solve all the mentioned problems, we think it is justified by its success. It
also provides some new physical principles, just like one had to introduce the notion of {\it relative} simultaneity in
the SR case, we have a new notion of space-time and then of course simultaneity. The reference frames are also not only
described by their speed, but also by their mass, or momentum. One shouldn't forget indeed that DSR should be seen as a
flat limit of QG.

First we give a physical interpretation of the coproduct that was thought to be describing the scattering. In short it
describes the way of how the momentum is changing when changing of reference frames. Following this new physical
principle, one can define the correct notion of scattering, associated with the symmetry inherent to the new regime,
the Poincar\'e-de Sitter symmetry. One needs to consider the pentamomentum. There is then a new  notion of energy, just
as when moving to the relativistic regime one gets the relativistic energy, we have now the DSR energy.  In this
scheme, the soccer ball problem is naturally solved as we show. Note that our scheme is giving a more rigorous approach
to the Magueijo-Smolin trick. It is natural to see then some new notion of space-time arising from this new symmetry.
We describe it quickly. There are in fact two aspects of it that can be seen by an heuristic argument equivalent.
Finally, We recall then what is the correct way to deal with the notion of multitude of deformations.

\subsection{Reference frames have a speed and a mass}
In this section we show how the  problem of the interpretation of the non commutativity or non associativity can be
solved by introducing a a refinement of the description of a reference frame.

First let us comment however that these features shouldn't be too worrying. Indeed, Special Relativity is an example
where the the non commutativity and the non associativity are a physical feature, even measured. Let us get into more
details.

When adding speeds which are not collinear,  the addition $v_1\oplus v_2$ is non commutative and non associative.
$$
v_1\oplus v_2 = Ad_h(v_2\oplus v_1)\neq v_2\oplus v_1 ,
$$
where $Ad_h$ is the adjoint action of $SO(3)$. The non commutativity is coded in the Thomas precession, which is even
present at speed much smaller than the speed of light. This precession comes from the fact that we are using the
hyperboloid structure to add the speed. More precisely, the speeds are parameterized by the boosts
($g_i=e^{f(|v_i|)\overrightarrow{v_i}.\overrightarrow{\sigma}}$, $\sigma_i$ Pauli matrices), and the addition of two
speeds is given by the product of two boosts, which is a rotation (the precession,
$R=e^{i\theta(v_i)\overrightarrow{n}(v_i).\overrightarrow{\sigma} }$) times a boost $g$. This structure comes from the
coset structure  ($SO(3,1)/SO(3)$) underlying the theory.
\begin{equation}\label{coset1}
g(v_1).g(v_2)=R(v_1,v_2).g(v_1\oplus v_2).
\end{equation}
This non commutativity is naturally implying a non associativity, and from this example one sees that non-associativity
is, contrary to the common sense,  physical, and    even a physical evidence of a new regime!\footnote{Note however
that this non associativity is deformation dependent, as there exist some which although non commutative are still
associative (co-associative coproduct), i.e. bicrossproduct basis like.}

Note that non commutativity is precisely encoding the {\it relative} point of view. We have a particle $Q_1$ with speed
$v_1$, with respect to the reference frame $R1$ (e.g. another particle $Q_2$). This latter has in turn a speed $v_2$
with respect to a reference frame $R2$. If one asks the question what is the speed of $Q_1$ in $R_2$, it is natural to
have $ v_1\oplus v_2$, there is a natural order, which encodes this idea of relativity\footnote{There is a clear
distinction between the composition of the speeds between different reference frames, i.e. their relative speeds, and
the scattering of the particles. We shall come back to this point in the next section.}.

Following the problem soccer and this example, we take two lessons: first the multiparticle states shouldn't be
described by the deformed coproduct. The non commutativity (or non associativity) can be interpreted as a hint of
relativity.

Before getting to the key physical interpretation let us point out that  in the Snyder construction we encounters the
same structure as in the Thomas precession. We shall come back to this into more details in the  section \ref{pheno}
but let us present how non associativity  could be present. Note again that the non associativity found in the context
of the Magueijo-Smolin trick has a different origin, and will be explained in the section \ref{solsoccer}.

Consider a Snyder like deformation. We have that the momentum is given by the term of the de Sitter boost $g=
e^{if(|p|) p^{\mu}J_{\mu4}}$, and from the coset structure of the de Sitter space we have that
\begin{equation}\label{coset2}
g(p_1).g(p_2)= L(p_1, p_2)g (p_1\oplus p_2),
\end{equation}
where $L$ is the Lorentz transformation depending on the momenta $p_i$. This encodes the addition of momenta in the
Snyder model. It is not difficult to be convinced that this addition is both non commutative and non associative.

Following the SR example we propose the following postulate.

\medskip
{\it Scattering of particles in the DSR regime is not described by the sum of the quadrimomenta, but by the sum of
pentamomenta, associated with the new symmetry, $SO(4,1)$. The deformed  addition of quadrimomenta is describing how
the momentum of a particle is changed by changing of reference frame. The notion of non commutativity  is precisely
encoding the notion of relativity. A non associativity, such as found in the Snyder case, encodes a notion of
precession (Lorentz).}

\medskip
 To summarize, we have now the particle described by the triplet
$(\overrightarrow{v}, p^{\mu}, \pi^A)$. The pentamomentum $\pi^A$ is described in the following section
\ref{pentamomentum}.

Our postulate should appear natural. In the context of DSR, which by construction is containing some gravity effects,
one should  expect that the mass associated to the reference frame matters. Then it is natural to say that a reference
frame (RF) will be defined by a speed and a mass  which is equivalent to say that is it defined by  a speed and a
momentum.

We want now to give an illustration to the fact that the  momentum of a RF is relevant, just as much as  its speed, and
for this we consider the simple case where there is no SR effect, nor DSR. Consider  a (quantum) particle ($Q$) of mass
$m$. We want to localize the particle, so this is done with respect to some reference frame\footnote{We are dealing
with a 1d model to give the general argument, for more details see \cite{aharonov}. } (e.g. another quantum particle),
which we note RF1. Assume now that RF1 is moving with speed $v_1$ with respect to another reference frame RF2. The two
reference frames are described by an origin and a mass $(O_i, M_i)$, $i=1,2$, and $O_1$ has for coordinate $x_1$  in
RF2 (equivalently, $O_2$ has the coordinate $-x_1$ with respect to RF1).
$$
\begin{array}{c|c|c}
& \rm{RF1} & \rm{RF2} \\\hline \rm{particle}& (x,v)& (x-x_1, v+v_1)\\
\hline \rm{RF1} && (x_1, v_1)
\end{array}
$$

As we are in a quantum case, we can assume that we can have a sharp measurement on $v$ (but then not on $x$). On the
other hand, we know that in a RF, one cannot determine its state of motion. This implies that at the same time we can
have a sharp measurement on the position of the center of mass of the system RF1 + particle, $x_{cm}$. In the quantum
mechanical setting, this means that
\begin{equation}\label{cmv}
[x_{cm}, v]=0.
\end{equation}
With a bit of algebra, using that $v=\frac{p}{m}$, $[x,p]=i$, $x_{cm}=\frac{M_1x_1+m(x-x_1)}{m+M_1}$, one can see that
this relation is not fulfilled but equal to $\frac{i}{m+M_1}$. This is because we are not considering the effect of
feedback of the frame on the particle and in this way we are not considering the right momentum: it must take into
account some sort of quantum potential $A$ linking the particle with the frame. One defines then the new momentum by
\begin{equation}
mv=p+mA,
\end{equation}
and we want to determine $A$. Using the commutation relations (\ref{cmv}) in the two RF, one gets at the end that
\begin{equation}
A=\frac{p+p_{O_2}}{M_1} +C.
\end{equation}
We see therefore that the "good" momentum of the particle is now defined as $p=\frac{M_1}{m+M_1}(mv-\frac{m}{M_1}p_2)$.
The right notion of momentum involves therefore the mass of the RF. It is also interesting to note that even if RF1 is
not in motion with respect to RF2, there is a non trivial contribution of the mass of RF1 in momentum of the particle
$p=\frac{M_1}{m+M_1}mv $. In the usual setting one considers that the mass of the RF is infinite (e.g. the RF is
"classical" contrary to the quantum particle), in which limit one recovers the usual momentum $p=mv$. In this limiting
case, the only relevant quantity is then the relative speeds of the reference frames. In a gravitational context, one
cannot put masses at infinity, at ease. This is why it is important to consider the mass of the reference frame and
then that it makes sense to talk about composition of momenta, which is of course {\it different than scattering}.

This is a key point we want to make in order to understand this new structure arising in  DSR. When defining a frame
one needs to define its relative motion with respect to another one, but also its mass. A DSR particle then will have a
momentum which takes into account  the momentum of the RF. The momentum of a particle is then also relative to the
reference frame and it makes sense  to define a notion of composition of momenta of reference frames.

 Special Relativity has introduced a deformation of the change of reference frame for the speeds so that
the total speed is bounded by $c$. Deformed Special Relativity is going now to introduce a deformation of the
composition of momenta which describes the change of momentum under a change of reference frame. It is not only the
speed which is relative to a RF but the momentum as well. The new composition \footnote{We insist (again!) on the word
"composition" opposed to the term "scattering".} for momenta is now given by the formula which precisely encodes the
relativity by this non-commutativity (or non associativity). Just as in SR, if one considers three frames 1, 2, 3, with
respective momentum $p_i$ then the deformed addition $p_1\oplus p_2$ is the  momentum of the RF 2 seen from the RF 1 in
the RF 3, whereas $p_2\oplus p_1$ is the resulting momentum the  momentum of the RF 1 seen from the RF 2 in the RF 3.
The two final momenta are related by a Lorentz transformation.

As the underlying space is non commutative, due to this deformed composition, one sees a {\it relative non commutative
geometry} \cite{relatncg}, that is the non commutative space-time we reconstruct for DSR is relative to a reference
frame or an observer.

As a last comment, we could argue that we used to illustrate a quantum particle, although DSR is classical. This
allegation is not exactly true as one is considering phenomena which involve the Planck length or others, and in any
case the Planck constant. In 4d, we expect that $\hbar\sim G$, i.e. the quantum effects comparable to the gravitational
ones,  so that $l_P\dr0$, but $M_P$ stays fixed. In this sense one is having a effective approximation of the quantum
gravity effects, and to deal with a quantum particle is perfectly adequate as an illustration.

To sum up, the spectator problem disappear now as we see that the non commutativity precisely encodes that there is a
notion of relativity involved and moreover, the momenta scattering   is not the one given by the coset structure which
describes the momentum composition. This is the same thing that occurs in the Special Relativity context, there is no
spectator problem as we have a relative composition of speeds and  the scattering is not given by the galilean momenta
but by the relativistic momenta for which the scattering is trivial.

\subsection{DSR momentum, and DSR energy}\label{DSRmomentum}

As we mentioned before, in the DSR regime, we expect a change of the symmetry group. One replaces the Poincar\'e group
$ISO(3,1)$ by the Poincar\'e de Sitter group $ISO(4,1)$, just like one changed of symmetry when moving from the
galilean to the relativistic case. The new momentum that we want to consider is given in terms of the translations on
the 5d Minkowski space, which we note $\pi^A$. One can also say that they carry the representation of the de Sitter
group. The Casimir of $ISO(4,1)$ is now constrained to be
\begin{equation}
\pi^A\pi_A=-\kappa^2.
\end{equation}
$\kappa$ has still dimension of a mass, and the pentamomentum $\pi^A$ is a function the quadrimomentum $p^{\mu}$,
defined by the adequate deformation, see section \ref{soldeformation}. In this sense we are dealing with a constrained
representation of $ISO(4,1)$. We shall see in the section \ref{spacetime}, the new notion of space-time coming with
this pentamomentum. However before this, one can try to give a physical interpretation to this new momentum.

According to the different basis one chooses, we shall get different expressions of the quadrivector momentum in terms
of the pentavector. In the Snyder case, this expression is particularly simple, and it gets more difficult in the other
basis. Note however that we can deal with this pentavector  independently of the chosen basis, it will be only at the
end when trying to extract physics that the choice of basis matters. We shall argue about these different choices of
basis, and why actually Snyder's seems to be a more natural choice, in section \ref{soldeformation}. In any case, by
introducing this new pentamomentum we expect some new physics to appear!

Let us choose the Snyder deformation as it is so close to the SR case, we can follow the same line as developed in
\cite{SR}. Again, the other deformations can also be obtained from this case, by an adequate transformation. We have
therefore
\begin{equation}\begin{array}{rcl}
c\kappa\pi^{\mu}&=&p^{\mu}\pi^4\\
\pi_4^2&=&\frac{\kappa^2}{1-\frac{1}{(c\kappa)^2}p^{\mu}p_{\mu}}=\kappa^2\Gamma^2.
\end{array}
\end{equation}
This factor $\Gamma$ is very close to the usual $\gamma^2=\frac{1}{1-\frac{1}{c^2}v^2}$, that one finds in the SR case,
which explains the notation. We have then that $c\pi^{\mu}=\Gamma p^{\mu}$, very similar to the SR case! Note that in
the chosen metric $(+----)$, the quadrimomentum of a particle is time-like and so we have indeed $0<\Gamma<\infty$, and
the quadrimomentum is bounded by the mass $\kappa^2$.

As first features of this DSR momentum, notice that $\pi^A$ is not bounded, and that when $ p^{\mu}p_{\mu}\ll
\kappa^2$, the effective momentum coincides with the usual one, i.e. we recover the relativistic regime. This is the
notion of classical limit we are looking for. On the other hand if one considers $ |p^{\mu}p_{\mu}|\sim \kappa^2$, it
becomes divergent, we are fully in the DSR regime.

In the SR case, with respect to the galilean point of view one interprets the $\gamma$ as giving the new notion of
energy, the relativistic energy. It is very tempting to interpret then $\Gamma$ as the new concept of energy, DSR
energy.

Before getting to this let us recall the usual case. The relativistic  energy is just given as $E^2-p_ip^ic^2=m^2c^4$,
from the Poincar\'e Casimir. Momentum generates translation in space, while the energy, considered as the Hamiltonian,
generates the motion of the particle/system, so that in the usual relativistic case one has,
$$ \{p^{i}, x^{j} \}=\eta^{ij}, \;\; \{p^0, x^i\}=\frac{p^i}{m}.$$
Just as we moved from the galilean energy (and 3d momentum) to a relativistic energy (and 3d momentum), one moves to a
new notion of energy in the DSR regime. Indeed in this new regime, the previous relations are deformed, i.e. non linear
anymore.
\begin{equation}
\begin{array}{rcl}
[X_j, p_i]&=& i\hbar(\delta_{ij}-\f{p_ip_j}{(c\kappa)^2}),\\
{[}X_i, p_0{]}&=& i\hbar\f{p_0p_j}{(c\kappa)^2}
\end{array}
\end{equation}
The fact that they are not linear anymore means that for example  this notion of relativistic energy in the DSR context
in not the correct one. The reason why linearity is important is when considering composite systems. Indeed for {\it
free systems},  energy and momentum are assumed to be {\it extensive quantities}. This generically requires linearity
of the bracket between position coordinates and momentum. The relativistic quadrivector is not extensive anymore, this
is the  manifestation of the soccer ball problem. This  means in particular that the relativistic energy is not the
good notion of energy anymore, it fails to take into account the interaction with the reference frame. We propose
therefore to introduce $\kappa \Gamma=\pi^4$ as the DSR energy\footnote{This has to be compared with the change
$E_g=\frac{p^2}{2m}\rightarrow E_r^2-p^2=m^2c^4$. As a direct feature of this change we have the notion of rest mass $
E_g=0 \dr E_r= mc^2$.} $\Ee=\pi^4c^2$. We have then that
\begin{equation}
-\Ee^2 + \pi^{\mu}\pi_{\mu}c^4=-\kappa^2c^4.
\end{equation}
This is a new kind of dispersion relation. The new feature that it implies on the relativistic dispersion relation will
be developed in the next subsection.  In the meanwhile, let us introduce the notion of rest DSR energy. It occurs when
one deals with a massless particle. In the usual relativistic interpretation, there is no rest energy, but it is easy
to see that it shall have a  rest DSR energy
\begin{equation}
\Ee^2=\kappa^2c^4.
\end{equation}
It is very important to understand that the DSR energy is essentially different than the relativistic energy. It
precisely encodes the shift between the relativistic energy and the interaction with the reference frame.  It is not
difficult to see also that it  makes linear the previous commutators
\begin{equation}\label{snyder2}
\begin{array}{rcl}
{[}X_i, \Ee{]}&=&\frac{ic^2}{\kappa}\pi_i,\\
{[}X_i, \pi_j{]}&=& -\frac{i}{\kappa}\delta_{ij}\pi_4, \\ \; [X_i, \pi_0]= 0, \; {[}X_0, \pi_4{]}&=&
\frac{i}{\kappa}\pi_0, \; [X_0, \pi_0]= \frac{i}{\kappa}\pi_4,  \; [X_0, \pi_i]=0.
\end{array}
\end{equation}
These relations are linear and define a Lie algebra structure. In the section \ref{solsoccer}, we shall show it allows
also to define  extensive quantities.



As a conclusion, we are proposing  a shift from the Poincar\'e group to the de Sitter-Poincar\'e group, in the same way
we moved from the Galilean group to Poincar\'e group. When this last move was accomplished it was accompanied by a new
conception of the notion of energy, this is the famous Einstein's relation, $E=mc^2$, for a particle at rest (more
generally this is the Poincar\'e Casimir $p_{\mu}p^{\mu}=m^2$). This new notion of energy was essential to understand
the nuclear structure of molecules and atoms. Indeed the notion of binding energy is represented either as an energy or
a (defect or excess of) mass. We expect the same kind of situation to happen, we explore this in the next section.

\subsection{The soccer ball problem} \label{solsoccer}
In this section we use the pentamomentum to describe the notion of "classical limit", in this way we solve the "soccer
ball problem". We gave a clear argument  in the previous subsection to interpret the deformed addition of quadrivector
momentum not as scattering but as  composition of the momenta of different reference frames. We show here how to define
this scattering. We first define what is in our sense the notion of classical limit, and then show how the DSR momentum
is solving this classical limit problem (i.e. define the right notion of scattering). We introduce also at the end  the
notion of binding DSR energy, which we mentioned heuristically in the section \ref{minmax}.

As mentioned in the section \ref{minmax}, we start by defining an observer, having a resolution $\delta l$, which is
looking at a system having a characteristic scale of $L$. In general, we have that $L_P\leq\delta l\leq L$. The maximal
mass contained in the system is
$$
\kappa\sim \frac{\hbar}{c L_P}\frac{L}{\delta l}.
$$
However in the DSR context we have that $\delta l \sim L \sim L_P$. We want then go from the Planck scale $L\sim L_P$
to a macroscopic scale $L\gg L_P$. As we have also for the DSR case, $\delta l \sim L_P$, this is equivalent to $L\gg
\delta l$, i.e. $\frac{L}{\delta l}\gg 1$, or $\frac{\delta l}{L}\rightarrow 0$. This is the notion of classicality we
expect and this can therefore be achieved in two equivalent ways:
\begin{eqnarray}
&& \textrm{fixed } L,\;\;  \delta l \rightarrow 0 \label{dl}\\
 && L\rightarrow \infty,\;\; \textrm{fixed } \delta l.\label{L}
\end{eqnarray}
The usual notion of classicality is reproduced by the first case. This is analogous for example to $\hbar\dr 0$ in
quantum mechanics. The second one  is more related to the notion of renormalization. Indeed one keeps a microscopic
structure fixed in order to study the large scale structure, this is the idea of coarse-graining.

More precisely, starting off with a system with length scale $L$, one should cut it up in (space-time) regions of size
$L_P$, each governed by DSR and eventually interacting with each other. Taking many copies of the same original
$\kappa$-Poincar\'e algebra and extracting from it the \lq\lq coarse-grained" deformed Poincar\'e algebra, we should
find a  larger $\kappa$ parameter.

In fact we see that by taking the for example two such systems of size $L_P$, then the associated resolution  should be
of the type $\frac{L_P+L_P}{2}=L_P$, i.e.  fixed. However this transformation is equivalent to having a characteristic
scale $L'$, twice as big as before, $L'=2L$ and by consequence   the mass is going to change linearly
\begin{equation}
\kappa'\sim \frac{\hbar L}{cL_P^2}+ \frac{\hbar L}{cL_P^2}=\frac{2\hbar L}{cL_P^2}= 2\kappa.
\end{equation}
In this sense, by taking many pieces, we have then naturally $\kappa\dr \infty$. We should now implement this idea
using the algebraic structure. This notion of classical limit is of course consistent with $\Gamma\dr0$.

\medskip

Let us remind what is the natural construction of the scattering  in Special Relativity from the Poincar\'e algebra
point of view. One considers two Poincar\'e algebras $\ppp_i$, $i=1,2$, with generators $J_i, P_i$ the Lorentz
transformations and the translations. One then expects the new state to be described by another Poincar\'e algebra. The
scattering is described by a trivial coproduct which corresponds in fact to taking the diagonal algebra. More
precisely, this is the algebra given by
\begin{equation}
\begin{array}{rcl}
P&=&p_1+p_2 ,\\
J&=& J_1+J_2.
\end{array}
\end{equation}
The two algebras are commuting with each other $[\ppp_1,\ppp_2]=0$. This new state generates a new redefinition of the
coordinates, as indeed one has that $[p_i^{\mu}, x_i^{\nu}]=\eta^{\mu\nu}$, this implies that we have the new position
operator $X=\frac{x_1+x_2}{2}$ so that we have $[P^{\mu},X^{\nu}]=\eta^{\mu\nu}$. An important point is  that there is
a compatibility between the coproduct associated to the change of reference frames and addition of the energy-momentum
for the scattering: the law of scattering is compatible with the relativity principle (the reader can find more details
on this in \cite{SR}).

The given construction here is completely algebraic and is independent in some sense of the particle. Indeed, a
particle is given as a representation of the Poincar\'e group  (mass and spin). If one wants to be more precise, one
can go to the algebra $\ppp_{cm}$ of the center of mass of the two particles which is then constructed from the
diagonal part of the product of the representations of the two algebras $\ppp_{m_i,s_i}$.
\begin{equation}
\ppp_{cm}\hookrightarrow \ppp_{m_1,s_1}\otimes\ppp_{m_2,s_2}.
\end{equation}
The coordinates are getting then pondered  by the masses. For a derivation of the calculation see for example
\cite{osborn}.

\medskip

We expect the same structure to apply to the DSR case. What one needs to consider is the diagonal algebra constructed
from the product of the two algebras of the de Sitter-Poincar\'e group generated by $(J^{AB}, \pi^A)$. In this case we
have therefore that
\begin{equation}\label{}
\begin{array}{rcl}
\pi&=&\pi_1+\pi_2 ,\\
\tilde J&=& \tilde J_1+\tilde J_2.
\end{array}
\end{equation}
The average position operator $X'=\frac{X_1+X_2}{2}$ is acting on the new momentum as
\begin{equation}\label{snyder3}
\begin{array}{rcl}
[X_i', \pi_j]&=& -\frac{i}{2\kappa}\delta_{ij}\pi_4,  \; [X_i', \pi_0]= 0, \; [X_i', \pi_4]=\frac{i}{2\kappa}\pi_i, \\
{[}X_0', \pi_4{]}&=& \frac{i}{2\kappa}\pi_0,  \; [X_0', \pi_0]= \frac{i}{2\kappa}\pi_4,  \; [X_0', \pi_i]=0,
\end{array}
\end{equation}
There are two comments to make at this point, the first concerns the behavior of $\kappa$, and the second the
representation of $X'$ in terms of the quadrivector $p$.

We have indeed that (\ref{snyder3}) is naturally implying a rescaling of $\kappa$
\begin{equation}
\kappa\rightarrow \kappa'=2\kappa.
\end{equation}
 This rescaling  is the expected one from the many particles picture. In this sense we are
seeing that this effective momentum is naturally solving the classical limit problem. Note that we could ask about the
center of mass for $\pi$. This is a much harder question than the relativistic case, as now space-time becomes a non
commutative space, and therefore the notion of center of mass is much more intricate and requires a deep study. We
leave this point for further investigations.

The factor $\frac{1}{2}$ for the position operator is a natural factor when considering the average of the position,
but it could be seen also as necessary as  in the non deformed  Poincar\'e case. Indeed when expressing the $\pi$ in
terms of the quadrivector momentum $p$ (that is choose a basis), one gets the typical relation between $p$ and
position,
\begin{equation}
[X^{\nu}, p^{\mu}]= \eta^{\mu\nu}+ O(\frac{p^2}{\kappa^2}),
\end{equation}
which is of course necessary if one wants to retrieve the classical Poincar\'e algebra when $\kappa\rightarrow\infty$.

 The next remark concerns the representation of $X'$, when considering a specific coordinate system. For example we
choose the Snyder coordinates. There is a non linear relation between the $\pi $ and the $p$, so there is a natural non
linearity for the position as dual of the $p$. One can see indeed that we have
$$p^{j}_{tot}=\kappa '\frac{\pi_1^j+\pi_2^j}{\pi_1^4+\pi_2^4}\neq p_1+p_2=\kappa\,\left( \frac{\pi_1^j}{\pi_1^4}+\frac{\pi_2^j}{\pi_2^4}\right). $$
This implies that the new representation of the space-time coordinates is
$$\begin{array}{rccl}
X^i=i \frac{\hbar}{c\kappa}(\pi^4\partial_{\pi^i}-\pi^i\partial_{\pi^4}) \dr X_i'= i \frac{\hbar}{c\kappa'}((\pi^4_1+\pi^4_2)\partial_{(\pi^i_1+\pi^i_2)}- (\pi^i_1+\pi^i_2)\partial_{(\pi^4_1+\pi^4_2)})\\
X^0=i \frac{\hbar}{c\kappa}(\pi^4\partial_{\pi^0}+ \pi^0\partial_{\pi^4}) \dr X_0'=i
\frac{\hbar}{c\kappa'}((\pi^4_1+\pi^4_2)\partial_{(\pi^0_1+\pi^0_2)}+ (\pi^0_1+\pi^0_2)\partial_{(\pi^4_1+\pi^4_2)})
\end{array}
$$

 This non linearity was in fact guessed by
Magueijo-Smolin in order to have a well defined classical limit. We can see that our construction is naturally
implementing this trick as also guessed in \cite{dsrgol}. In some sense this is natural as we are defining a trivial
addition and for the $\pi$, and get it back on the $p$, this fits exactly with their philosophy, but  we provide a more
exact construction.

Consider indeed the map $U$ from the 5d Minkowski space to the de Sitter space of curvature $\kappa$,
$$\begin{array}{rccl}U:&M^5&\rightarrow & de S_{\kappa}\\
&\pi^A &\rightarrow & p^{\mu}=\kappa\frac{\pi^{\mu}}{\pi^4}.\end{array}$$
 Then we can write $p_{tot}$ as
\begin{equation}\label{addition4d}
p_{tot}= U_{2\kappa}(U_{\kappa}\mone(p_1)+U_{\kappa}\mone(p_2))=U_{2\kappa}(\pi^1+\pi^2)=2\kappa
\frac{\pi_1+\pi_2}{\pi_1^4+\pi_2^4}=\kappa' \frac{\pi_1+\pi_2}{\pi_1^4+\pi_2^4},
\end{equation}
which we recognize as the  formula proposed by Magueijo and Smolin (of course the reader noticed we still used the
Snyder deformation).

Before moving to the problem of change of reference frame, let us consider the energy component of total quadrimomentum
in (\ref{addition4d}). We have
\begin{equation}
E_{tot}=c^2\,\left(\frac{E_1}{\sqrt{1-(m_1c\kappa)^2}}+
\frac{E_2}{\sqrt{1-(m_2c\kappa)^2}}\right)\left(\frac{1}{\sqrt{1-(m_1c\kappa)^2}}+ \frac{1}{\sqrt{1-(m_2c\kappa)^2}}\right)\mone.
\end{equation}
It is rather obvious that $\Delta E\equiv E_{tot}-(E_{1}+E_{2})$ does not vanish, and we can interpret this difference
as an interaction potential $V\equiv \Delta E$ between the two systems depending on their momenta. This potential would
forbid the momentum to exceed the bound $\kappa$.

\subsection{Change of reference frame for the pentamomentum}
 One of the basic postulates of physics is
the principle of relativity stating that different observers should still experiment the same laws of physics. More
precisely, we require to have the same laws of conservation in any reference frames. More technically this means that
we want a compatibility relation between the coproduct describing the scattering process (i.e going from two systems to
one coarse-grained composite system) and the coproduct describing the change of reference frame. More precisely
consider two systems with respective pentamomentum $\pi_1, \pi_2$, and relativistic momentum $p_1,p_2$ in a given
reference frame. The system $1+2$ has a total pentamomentum $\pi_1(p_1)+ \pi_2(p_2)$, which is conserved. Now assume
that this reference frame has a momentum $p_{\mu}$ with respect to another reference frame then the total pentavector
must be conserved in the new reference frame $\pi_1(p_1\oplus p)+ \pi_2(p_2\oplus p)$. The addition $p_i\oplus p$ is
given in term of the deformation and the formula (\ref{coset2}), which we restate for clarity.
\begin{equation}\label{coset3}
g(B_1)g(B_2)= L g (B_1\oplus B_2),
\end{equation}
where $L$ is the Lorentz transformation. Note that here we didn't introduce the momentum, which is related to the
quadrivector $B$ through the deformation map $f$. \be c\kappa f(\eta)B^{\mu}=p^{\mu}.\ee We come back to this relation
in the section \ref{soldeformation}. Note also that this structure is very similar to the relativistic case \cite{SR}
where in this case one had $L$ to be a rotation, and physically appeared as the Thomas precession. Here $L$ is a
Lorentz transformation and in this case we have a {\it Lorentz precession}. We come back to this in the section
\ref{pheno}. In any case, one can calculate the deformed addition, in the same way that this is calculated in the SR
case. It is rather obvious that this addition is both non commutative and non associative. This shouldn't be taken as a
problem but at the contrary as a new physical evidence of a new regime, just as the Thomas precession was a physical
evidence of the relativistic regime.

We want to show that under change of reference frame, the pentamomentum is still conserved. We shall show this for the
DSR energy, the rest following in the same way. In the appendix \ref{additionmom}, we show that for a given momentum
$p= c\kappa f(\eta)B^{\mu}$, and $g=e^{\eta iB^{\mu}J_{4\mu}}$, the DSR energy can be written
\begin{equation} \Ee=\t [V^\dagger g^\dagger V g]=\cosh\eta,
\end{equation}
with $V=\gamma_4\gamma_0$. Using the geometric addition of the momenta, we have that the DSR energy $\Ee_i$ of the
system $i$, after change of reference frame is given by
\begin{equation}
\Ee_i(p_i\oplus p)=\t [V^\dagger(g_ig)^\dagger V g_ig]
\end{equation}
After calculations indicated in the appendix \ref{additionmom}, we get that
\begin{equation}
\Ee_i(p_i\oplus p)= \cosh\eta_i \cosh\eta - \sinh\eta_i\sinh\eta_2B^i_{\mu}B^{\mu},
\end{equation}
so it is then easy to see that the total energy is conserved under change of reference frame
\begin{equation}
\Ee_{tot}'= \Ee_{tot}\cosh\eta - \sinh\eta \pi_{tot}^{\mu} B_{\mu}.
\end{equation}
Note that the order in which one composes the momenta is very important here. It precisely encodes the "order" of the
relativity.

Note also that we gave a proof of invariance for {\it all} homogenous deformations. It is pretty easy to check that the
 deformations given by the quantum group like deformations do not transform well under the change of reference frame.
In fact the deformations which give $\kappa$-Minkowski like space-times are not behaving well under the change of
reference frame. This can be interpreted from the fact that one has chosen a preferred direction (in general a time
direction), which then breaks the relativity principle. An possible interpretation is that the homogenous deformation
correspond to an inertial observer whereas the bicrossproduct-like deformations correspond to a accelerated observer.
This has however to be shown rigorously. In any case it is interesting that we have exactly the same situation in the
SR case, a bicrossproduct like basis \cite{SR}, and the same analysis should apply, or even be easier to do. The good
deformation, i.e. for an inertial observer is an homogenous one in this case. We leave this for further investigations.

\subsection{New concepts of space-time}\label{spacetime}

One of the biggest issue and obstacle in DSR is the reconstruction of space-time. Indeed the theory is defined by  its
energy-momentum sector and its algebra of symmetries. Thinking of DSR as a non-commutative geometry implementing a
$\kappa$-deformed Poincar\'e invariance, one introduces some operators corresponding to the space-time coordinate
sector. However, there is a lot of ambiguity in understanding the physical meaning of these coordinate operators.
First, there is a lot of freedom in defining such coordinate operators, and we do not know whether these different
choices define different DSR theories or simply express different basis for the same space-time. Then, coordinate
operators are not classical coordinate, so we still need how to recover a (semi-)classical notion of space-time notions
or equivalently a notion of fundamental events. Only solving these issues, we could talk about the space-time of the
DSR theory.

There have actually been many studies on analyzing the space-time sector of DSR. Besides the main approach of building
the space-time coordinate operators as the algebraic dual of the momentum space in a non-commutative geometry, one can
also find a proposal of inducing DSR through a canonical transformation on the phase space which automatically provides
us with the space-time sector \cite{phasespace}, or a detailed analysis of the definition of space-time events as POVM
(positive-operator-valued-measure) using the tool of induced representations \cite{toller}. However these analysis
always bump into strange effects. In the canonical transformation approach, the effective DSR metric depends explicitly
on the mass (of the observer) \cite{phasespace}. The POVM analysis derives an usual transformation law of the
coordinates under space-time translation such that it explicitly depends on the energy-momentum of the object defining
the event, which is interpreted as "a failure of the absolute character of the concept of space-time coincidence"
\cite{toller}. These results point toward the necessity of to take into account the mass of reference frame when
working in a DSR theory.

\medskip

From our point of view, DSR needs an extension of the principle of relativity for the four-dimensional space-time. A
reference frame is not only defined through its speed as in Special Relativity but now also a mass is attached to each
reference frame and allows to distinguish them. In particular, this leads to a new subtlety in the definition of
simultaneity: even at the same speed, two observers would define different slices of simultaneity (what they call
space) depending on their masses. This is similar to effects of General Relativity where masses curve space-time. This
is why it can be useful to see DSR as a theory halfway between Special Relativity and General Relativity trying to take
into account relevant effects of the mass of objects while staying in a flat context. Indeed, it was already shown in
\cite{dsrgol} that mapping a low-curvature space-time onto a flat space-time leads to describe it effectively through a
DSR theory. In the non-commutative geometry formalism of DSR, this extra dependence of the notion of simultaneity on
the mass of the reference frames is translated into a notion of {\it fuzzy simultaneity}. Indeed the time coordinate
operator doesn't commute with space coordinates, and it is not straightforward anymore to define slices of space. An
interesting project would thus be to construct semi-classical observers, in the non-commutative scheme, and consider
their corresponding slices of simultaneity. One would expect that, given a classical speed of the observer, one would
get a one-parameter ambiguity in the definition of the states of the observer which should correspond to its attached
mass, or to its scale factor.

\medskip

One can push further the study of the mathematical framework of DSR and the construction of coherent states in order to
possibly identify the scale factor in the definition of physical states. Since the theory is defined with some
space-time coordinate operators acting on a Hilbert space ($L^2$ functions on the de Sitter or anti de Sitter momentum
space), one can try to diagonalize the coordinate operators and study the average values and spread of the different
states. Ultimately, one would like to have coherent states representing the concept of semi-classical space-time points
\cite{dsrol}, or even generalize the notion of states to POVM as in \cite{toller}. A nice and easy-to-deal-with basis
of the DSR Hilbert space is given by states diagonalizing the de Sitter Casimir operator and the Lorentz Casimir
operator and thus are labelled by (irreducible) representations of the de Sitter group and its Lorentz subgroup. A
detailed analysis of the average values and spread of the coordinate operators will be found in \cite{cohstate}. It is
still possible to identify states peaked around any space-time coordinates, but the Casimir operator of the de Sitter
group provides us with an additional parameter describing the spread of the states around its mean value. Thus choosing
a certain resolution, one can write a system of coherent states covering the space-time and the Hilbert space of
states\footnote{Or alternatively, one can choose one point and consider all coherent states peaked around it. Allowing
all possible spread also provides a system of states covering the whole DSR Hilbert space.}. Such a state peaked around
a particular space-time point will be considered semi-classical for an observer whose resolution is larger than the
spread of the state. Thus a semi-classical DSR point is defined through its mean value and its spread, which are
independent: the definition of a DSR space-time point requires five numbers and the spread can be re-interpreted as the
length scale of observation or the resolution of the observer, i.e. a scale factor.

\medskip

In this section, we would also like to advocate using five space-time "coordinates" dually to the five dimensional
pentamomentum introduced previously. First, let us discuss the physical meaning of the fifth coordinate for the
energy-momentum. We are describing the momentum space as being the (Anti)de Sitter space as embedded in the
five-dimensional Minkowski space:
$$
-\kappa^2= \pi_0^2-\pi^2_1-\pi^2_2-\pi^2_3-\pi^2_4.
$$
At fixed $\kappa$, the momentum space is still a four-dimensional manifold. However, we have argued above that
renormalization is a normal and necessary feature of the DSR theory. Allowing the mass/energy bound $\kappa$ to be
renormalized solves the "soccer ball" problem. The parameter $\kappa$ then depends on the scale (of observation), and
changes accordingly to the relation between the mass and the length scale of a (Schwarzschild) black hole. Now, the
full theory is described in terms of (a deformation of) the usual notion four-dimensional energy-momentum plus the
bound $\kappa$ indicating the energy scale. We can equivalently describe the theory through the pentamomentum
$\pi_{0,..,4}$, which is actually a more symmetric object. Dual to this five-dimensional momentum, there should be a
notion of five-dimensional space-time. The extra space-time coordinate would have the interpretation of a scale, at
which the observer would be analyzing the space-time.

This proposal can be seen as related as the derivation of DSR theories at fixed $\kappa$ (usually the Planck mass)  as
gauge fixed versions of the conformal algebra (see for example \cite{conformalDSR}). There have also been some work
directly trying to realize the DSR algebra on a five-dimensional $M^4\times \R$ space-time i.e interpret DSR as a
five-dimensional "Galilean" relativity \cite{newtime}. In these approaches, one is usually lead to introduce a further
time variable $T$ different and independent from the standard time coordinate $t$ of the four-dimensional Minkowski
space-time. In \cite{newtime}, an analysis of particle dynamics in such framework shows that this new variable $T$ is
proportional to the proper time for the particle with the proportionality ratio depending explicitly on the mass $m$ of
the particle or more precisely on the ratio of $m$ to the maximal mass bound $\kappa$. $T$ can then be interpreted as a
time running at the "Planck scale" or more precisely at the maximal scale considered, while the normal time $t$ is
still the usual clock time running for the considered object. This way, assuming that the Planck scale time runs the
same for all objects, we get that the clock time flow depends on the clock mass, which is quite natural in the
framework of General Relativity.

Such considerations are very similar to the picture which arises from the study of the renormalization group flow of
General Relativity. Indeed, the logarithm of the cut-off scale can usually be interpreted as a time coordinate along
which the effective theory for general relativity gets renormalized. And it is actually possible to derive a
five-dimensional space-time representation for the renormalization group flow. As en example, \cite{percacci}
re-interprets the renormalization group equations in terms of the five-dimensional Anti de Sitter space-time of the
Randall-Sundrum model \cite{RS}.

A good framework to make this more explicit would be to develop the point of view of {\it unification}. The main idea
is that the introduction of a new dimensionful parameter $\kappa$ allows to introduce a further quantity to the
quadrimomentum and turn it into the pentamomentum, exactly in the same way that the introduction of the dimensionful
speed of light $c$ allowed to reunify the energy and the 3-momentum into a single object, the relativistic
quadrimomentum. Then, the whole question is the physical interpretation of the "new" fifth coordinate of the
pentamomentum. We propose to interpret it as a scale factor or scale of energy in a renormalization scheme. More
precisely, the equation of motion of this fifth component would reproduce a renormalization group equation. This
conjecture is currently under investigation.

\subsection{Deformations: one or many?}\label{soldeformation}
In this section we address the last problem we mentioned, the problem of the number of deformation.  A simple
restatement of the physical situation with the help of the  SR analogy is sufficient to solve it. A more geometrical
argument was also given in \cite{dsrgol}.

When dealing with the Special Relativity case \cite{SR}, the question of the number of physical deformation was clear:
once the physical quantities are determined, that is once we determined the basis, we can then do different rotations
of this basis, provided that we keep the the same physical entities.

More exactly, we have the deformation of the space of speeds, given by the map $g:\R^3\dr H$, where $H$ is the
hyperboloid. It codes how you embed the speeds in this new space. Once one has determined $g$, the coordinates on the
hyperboloid to define the speeds  are not physically relevant, as the physical quantities are not coordinate dependent,
but depend only on the map $g$.

One finally needs the experiments to tell us which is the right deformation we need to use. For example the contraction
of length (dilatation of time) or the Thomas precession provide us some good hint on the deformation. Note that a
priori it is only the experimental setting  which is giving us some information on the right deformation, there is no
abstract mathematical  argument.

We propose that the same analysis applies  to the DSR case as well. Now one has the map $f:\R^4\dr dS_{\kappa}$, which
encodes the deformation of the momentum space. The coordinate systems used on the de Sitter space are not physically
relevant, when the right physical notions are defined. The right deformation has to be determined by some experiments.
The current proposed ones   essentially aim on determining the deformed dispersion relations. In the next section we
shall provide a new predictions which should distinguish in particular the Snyder deformation to the other ones.

We proceeded by analogy but actually this argument is corroborated by the geometrical approach taken in \cite{dsrgol},
which shows how DSR can be seen as good approximation for Quantum Gravity.

Some general comments are however worth to do. Indeed DSR is really about deforming the composition of the momenta and
not the scattering. In this sense, we are introducing a new level in the Relativity principle. It is not only the
speeds that are relatives to frame but momenta as well. This principle is really constraining and naturally implies an
homogenous deformation. Indeed if a preferred direction is pinpointed, this would go against this principle. On a more
physical ground, it seems more natural for an inertial observer to have an isotropic point of view, which is not the
case for an accelerated observer, where at the contrary a preferred direction is naturally identified. In this sense
the non isotropic deformation could be conjectured to coincide with the accelerated observer, just as in the SR case,
where one can define the analogous of the $\kappa$-Minkowski coordinates \cite{SR}.

These two arguments tend to favor a deformation of the type
\begin{equation}
\begin{array}{rccl}
f:&\R^4&\dr&dS_{\kappa} \\
& p^{\mu}&\dr & e^{\kappa f(\eta)J^{4\mu}}
\end{array}
\end{equation}
with $p^{\mu}p_{\mu}= \kappa f(\eta)$, where $\eta$ is the angle of the de Sitter boost, $J^{4\mu}$.

The fact that  different observers are associated with deformations can be further developed. Indeed,   one can recall
that SR is invariant under passive diffeomorphisms, which are equivalent to the change of coordinates, whereas the
active diffeomorphisms are changing the physical quantities. In the same way DSR is invariant under the passive
diffeomorphisms and generalizing DSR so that it can be invariant also under the active diffeomorphisms would be a
theory equivalent to General Relativity for the DSR symmetry. In particular it uses a generalized equivalence
principle. See \cite{dsrgol} for more comments on this topic.

To conclude in a provocative manner, there is at the same time one and many physical deformation, the algebraic and
geometrical point of view coexist. The experimental input is now essential to fix what is the right deformation.

\section{Some new phenomena}\label{pheno}
This section is devoted to the first direct phenomenology that one can extract from this new framework. We just sketch
some of the new features, leaving the exact calculations for further investigations.

As the presentation of the new formalism went on, different aspects of a new phenomenology were clearly apparent. This
is natural as we are defining a new symmetry: a full new phenomenology should appear just as new phenomena appeared
when introducing SR.

This is of importance as the deformation has to be experimentally determined, and therefore one needs a couple of these
new phenomena to define it and then the others to check out it is consistent with the predictions.

Fortunately we are  providing here  two more physical situations as direct manifestations of the new symmetry. There
are some other physical situations which are usually proposed to be a manifestation of DSR, e.g. GZK cutoff, the
$\gamma$-ray bursts and so on. Those latter should be reconsidered in the context of a field theory expressed in terms
of representation of the new symmetry. It is only then that one can re examine the physical predictions relative to
these experiments, or the calibration of the deformation. This is work in progress.

\begin{itemize}

\item {\bf The Thomas precession}: As mentioned earlier, it indicates the relativistic regime, and is easily measurable.
It is constructed from the product of boosts. In order to respect the maximum quantity, one has to  realize them in a
non linear way. This non linearity should provide therefore some corrections to the calculations of the Thomas
precession. This is therefore a new experimental situation that one should explore in order to find evidences for DSR.
Note that this is situation is also found when implementing a maximal acceleration \cite{frederic}.

\item {\bf The Lorentz precession}: We saw that in order to add the quadrivector momenta, one has to consider also a Lorentz precession. This Lorentz
precession is not a  deformation of the Thomas precession, but a completely new object. It is however its analog as
coming from the quasigroup structure present in the de Sitter space. Given a resolution $\delta l$, consider three
particles, $P_1,Q_1, Q_2$  such that their mass is close $\kappa$. We are in the full DSR regime. The first particle
has a momentum $p_1$ with respect to $Q_1$, whereas $Q_1$ has a momentum $p_2$ with respect to $Q_3$. If one wants to
look at the particle $P_1$ in $Q_2$, then it will have a momentum $p_1\oplus p_2$. On the other hand if one wants to
look at the particle $Q_1$ with respect to $P_1$ in the reference frame given by $Q_2$, then it is obtained by acting
with a Lorentz transformation on the sum $p_1\oplus p_2$. More concrete physical situations where this could be
measured are to be still proposed.

\item  {\bf Varying speed of light}: We noticed in section \ref{minmax} that one could expect a variation of the speed of light,
by having a bound on mass. Essentially having a  resolution $\ell$, given by the wave length of the probing ray of
light, we cut space-time into cells of radius $\ell$.  We have a maximum mass-energy $M_\ell$ associated to each cell.
In these cells we have then some energy density $\delta M$ arising form the contained matter, (fields, ect...)  and the
cosmological constant (which can increase or decrease the total energy density). When $\delta M\ll M_\ell$, we are in
the usual relativistic scheme, whereas when $\delta M\sim M_\ell$ we are in the DSR regime. If we have for example
another ray of light with very high energy (so that $\delta M\sim M_\ell$) passing by in these  cells, then it will be
naturally slowed down (for an external observer). We obtain therefore a speed of light depending on the resolution
$\ell$ and the cosmological constant. This gives an heuristic way to link the variation of the speed of light with DSR.
Other arguments were already presented in e.g. \cite{vsl}. Note that the main arguments presented in the latter  are
developed in the context of a non linear realization of the Lorentz group and the Magueijo-Smolin trick based on the
renormalization group. As our framework contains these latter, the same arguments should also hold.
\end{itemize}

\section*{Conclusion}
In this article we proposed a new theoretical framework, inspired from Special Relativity, in order to solve the main
problems of Deformed Special Relativity. This framework consists in two main steps, first the interpretation of the
deformed addition of the relativistic momentum, not as a scattering, but as a composition of momenta, relative to some
reference frames. Reference frames are now not only described by their relative speeds, like in the SR case but also by
their mass. This seems natural if one has in mind that DSR should be an approximation of Quantum Gravity.  The second
steps is to change the symmetry, which are describing the DSR regime, and consider the pentamomentum associated to this
symmetry. We showed that this new momentum is naturally solving the soccer ball problem, in particular by reproducing
the Magueijo-Smolin's trick. As it was guessed in many occasions, the solution of this problem is really rooted in a
renormalization group idea. To summarize DSR can be interpreted as a new regime, just one had already before the
relativistic and galilean regimes. The key factor (not a constant!) here is $\kappa$ which indicates the maximal
mass/energy for a given scale. Finally, the problem of the number of deformations is also solved using a clear
geometric picture. We made clear that the deformation has a priori no chance to be determined by theory. It seems that
the experiments are the best way to get it.

 There is  a natural new  notion of space-time emerging from this new symmetry. It can be seen as a fuzzy space-time
with a resolution or an extended space-time with two times. It seems that heuristically these two concepts are related
by a renormalization group scheme. This link deserves however to be inquired more precisely.

An important point in this scheme is the interpretation of the fifth momentum coordinate, the DSR energy. In fact one
could argue that just as Special Relativity has unified the notions of space and time, DSR is going to unify the
notions of momentum space with the renormalization group scheme. Indeed, as we are dealing now with a momentum space of
five dimensions, one should study the equation of motion for the fifth coordinate, by defining some kind of action. The
conjecture would be that this equation of motion is the renormalization group equation.  This idea is currently under
studies.

Finally, with a new physical regime one expects to have some new phenomenology. On top of the few already existing
proposals among which the correction of the Thomas precession and the varying speed of light, we have introduced a
completely new physical manifestation: the Lorentz precession. It is important to have many of such experimental scheme
as some might be easier to test than other.

In the end, the main issue we are left with is how to couple the extended relativity principle to  the equivalence
principle and possibly obtain a deformed general relativity which would naturally take into account the Planck mass.

\section*{Acknowledgments}
We would like to thank Daniele Oriti for many discussions on the physical interpretation  of Deformed Special
Relativity and Laurent Freidel for discussions on the issue of unification in physics. We are also grateful to Lee
Smolin, Joao Magueijo and Jerzy Kowalski-Glikman, for their interest and encouragements on completing the present work.

\appendix

\section{Spinor Representation of the Lorentz and De Sitter groups}
The Pauli matrices are given by $$ \sigma_1=\left(\begin{array}{cc} 1& \\&-1
\end{array}\right), \; \sigma_2=\left(\begin{array}{cc} &i \\-i &
\end{array}\right),\; \sigma_3=\left(\begin{array}{cc} &1 \\1& \end{array}\right). $$
For simplicity we also note $\sigma_0=Id=\left(\begin{array}{cc} 1& \\&1 \end{array}\right).$

We introduce then the $\gamma_{\mu}$ such that
$$\gamma_0=\left(\begin{array}{cc} 0&\sigma_0\\-\sigma_0 &0\end{array}\right), \; \gamma_i=\left(\begin{array}{cc}
0&\sigma_i\\\sigma_i&0\end{array}\right),$$ and they satisfy both
\begin{equation}\label{gg1}
\begin{array}{c}
\{\gamma_{\mu},\gamma_{\nu} \} =2\eta_{\mu\nu}1_4, \;\; J_{\mu\nu}=\frac{i}{4}[\gamma_{\mu},\gamma_{\nu}],
\end{array}
\end{equation}
with $(-,+,+,+)$, for signature for $\eta_{\mu\nu}$.

\begin{equation} \gamma^{\mu}\gamma^{\nu} =\demi([\gamma^{\mu},\gamma^{\nu}
]+\{\gamma^{\mu},\gamma^{\nu}\})= \demi (2\eta^{\mu\nu}-4iJ^{\mu\nu})
\end{equation}

The Lorentz algebra is given by the $J_{\mu\nu}$, with $\mu,\nu=0\hdots3$,
$$ J_{ij}= -\demi \epsilon_{ij}^{k}\left(\begin{array}{cc} \sigma_k& \\&\sigma_k \end{array}\right)\; \;J_{0i}= \frac{i}{2}\left(\begin{array}{cc} \sigma_i& \\&-\sigma_i \end{array}\right) $$

We introduce now the de Sitter group, generated by the element $J^{AB}$, $A,B=0\hdots4$ which contains as a subgroup
the Lorentz group. To construct the de Sitter group, we consider the $\gamma^{\mu}$ and add the chirality operator,
\begin{equation}
 \gamma_4=\gamma_0\gamma_1\gamma_2\gamma_3=\left(\begin{array}{cc} -1&0 \\0&1\end{array}\right).
\end{equation}
The $\gamma^A$ satisfy then the analogous of (\ref{gg1}),
\begin{equation}\label{gg2}
\begin{array}{c}
\{\gamma^{A},\gamma^{B} \} =2\eta^{AB}1_4, \;\; J^{AB}=\frac{i}{4}[\gamma^{A},\gamma^{B}],
\end{array}
\end{equation}
where $\eta^{AB}$ is the 5d Minkowski metric, with signature $(-,+,+,+,+)$. The generators $J^{4\mu}$ are called the de
Sitter boosts generators, and take the form
\begin{equation}
J_{4k}= \frac{i}{2}\left(\begin{array}{cc} &-\sigma_k \\\sigma_k& \end{array}\right),\;\; J_{40}=
\frac{-i}{2}\left(\begin{array}{cc}&  \sigma_0 \\\sigma_0& \end{array}\right).
\end{equation}
In order to simplify the calculations, we give a useful parameterization of the  de Sitter boosts $e^{i\eta
B^{\mu}J_{4\mu}}$, where $\eta$ is the angle of the boost. We choose to have  $B$ to be time-like and normalized
$|B|=B^{\mu}B_{\mu}=-1$, we leave the case light-like for further investigations. We have also that
$$J_{4\mu}J_{4\mu}=\frac{1}{4}\eta_{\mu\nu}- \frac{i}{2}J_{\mu\nu},
$$ which allows then to get
\begin{equation}\label{des}
e^{\eta 1B_{\mu} J^{4\mu}}= \cosh\frac{\eta}{2}\one+ 2i\sin\frac{\eta}{2} B_{\mu} J^{4\mu}.
\end{equation}

\section{Geometric addition of quadri-momenta in DSR}\label{additionmom}
We can parameterize the de Sitter space using the de Sitter boosts.
Essentially, we have that (putting $\kappa=1$, and $g$ de Sitter boosts)
\begin{equation}
V^{\dagger}g^\dagger V g= \cosh\eta \one + 2i\sinh\eta B^{\mu} J_{4\mu},
\end{equation}
which can be identified with $\pi^4=\cosh\eta, \pi^{\mu}=\sinh\eta B^{\mu}$. $V$ is the origin of the de Sitter space
and $V=\gamma_4\gamma_0=V^{\dagger}$. The calculation follows easily once remarked that
\be\label{propJ1} \begin{array}{c}J_{4k}^\dagger = J_{4k}; \;  J_{40}^\dagger =- J_{40}\\
\gamma_4 J_{4\mu}= -J_{4\mu}\gamma_4\\
 \gamma_0 J_{4k}= J_{4k}\gamma_0; \;  \gamma_0 J_{40}= -J_{40}\gamma_0. \\  \end{array}\ee


It is easy to see that the DSR energy is easily extractable from this formula,
$$
\Ee= \t V^\dagger g^\dagger V g.
$$
As mentioned previously, to add momenta, one considers the geometric addition on the de Sitter space that is the
product of de Sitter boost.
\begin{equation}
g(B_1)g(B_2)= Lg (B_1\oplus B_2).
\end{equation}
Under change of reference frame, $p_i\dr p_i\oplus p$, given by $g_i.g$ the DSR energy behaves therefore as
\begin{equation}
\Ee_i= \t V^\dagger(g_ig)^\dagger V (g_ig).
\end{equation}
By using the properties (\ref{propJ1}), we get
\begin{equation}
\begin{array}{rcl}
\Ee_i&=& \t V^\dagger(g_ig)^\dagger V (g_ig) \\
&=&\t V^\dagger V(gg_i) (g_ig)\\
&=&\t g^2g_i^2,
\end{array}
\end{equation}
 and we obtain the final formula
\begin{equation}
\Ee(p_i\oplus p)= \cosh\eta_i \cosh\eta - \sinh\eta_i\sinh\eta_2B^i_{\mu}B^{\mu},
\end{equation}
where one can find the quadrivector momentum by using that $p^{\mu}= c\kappa f(\eta)B^{\mu}$.

\end{document}